\documentclass[journal]{IEEEtran}

\usepackage{amsmath,amssymb,amsfonts}
\usepackage[T1]{fontenc}
\usepackage[utf8]{inputenc}
\usepackage{graphicx}
\usepackage[export]{adjustbox}
\usepackage{booktabs}
\usepackage{multirow}
\usepackage{array}
\usepackage{float}
\usepackage{cite}
\usepackage{url}
\usepackage[unicode]{hyperref}
\graphicspath{{mathpix_assets/}}
\DeclareUnicodeCharacter{2713}{\checkmark}
\DeclareUnicodeCharacter{00D7}{\ensuremath{\times}}

\title{A Multi-Frequency Input-Admittance Model of Locomotive Rectifier Considering PWM Sideband Harmonic Coupling in Electrical Railways}
\author{Xiangyu Meng, Zhigang Liu, Guorong Li, Xunjun Chen, Siqi Wu, Keting Hu}

\begin{document}
\maketitle

\begin{abstract}
Electrical railway harmonic instability issues are common in the high-frequency range. The effective frequency of the traditional converter's small-signal averaging model is below 1/2 switching frequency since the pulse width modulation (PWM) sideband harmonic components are ignored. In this article, the dynamic propagations of perturbation frequency and the generated PWM sideband components are constructed first. Then the locomotive rectifier's multi-frequency input-admittance model is derived appropriately. Afterward, an admittance conversion approach is used to convert the multi-frequency model into the single-input-single-output (SISO) model whereas retaining the sideband frequency couplings. The proposed SISO model is more accurate than the traditional small-signal averaging model in the frequency range higher than $1 / 2$ switching frequency. It is found that PWM sideband harmonics dominate the locomotive rectifier's input-admittance characteristic higher than $\mathbf{1} / \mathbf{2}$ switching frequency. Finally, based on the proposed model, the influence of different switching frequencies, control bandwidths, and traction network impedance on system harmonic stability is revealed by the hardware-in-the-loop (HIL) results.
\end{abstract}

Index Terms-Electrical railway, harmonic instability, inputadmittance model, locomotive rectifier, StarSim.

\section*{I. Introduction}
WITH the rapid expansion of electrical railways in China, many new types of trains with ac-dc-ac traction drive systems, such as electric multiple units and high-power locomotives (collectively referred to as locomotives), have been put into service. Onboard converters, particularly rectifiers, are typical nonlinear equipment, resulting in com-

plicated interactions between the locomotives and traction network (L-N) system. As a result, interactions within the L-N system can lead to instability issues such as highorder harmonic resonance (HHR) [1] and harmonic instability phenomenon (HIS) [2]. These events will amplify the traction network voltage and current, endangering the safety and reliability of the L-N system. The HHR is mainly caused by the interaction between the intrinsic resonance sites of the traction network and the distinctive harmonics created by the locomotive rectifier switching process, which has been intensively investigated [2]. The magnitude of voltage and current in resonance frequency is constant when HHR occurs, indicating that the L-N system is critically stable. However, in comparison to HHR, the HIS is unstable, characterized by continuously oscillating on one or several frequencies [3]. The article examines the L-N system's instability as a result of HIS.

The impedance-based small-signal averaging approach is currently an effective tool for analyzing system stability issues [4]. The small-signal averaging model assumed that the pulse width modulation (PWM) process only contained a perturbation frequency because the averaging process in the low-frequency region allows the generated sideband harmonic components in the sampling and PWM processes to be ignored. As a result, the effective frequency range of the model is smaller than the $1 / 2$ switching frequency. However, the model is no longer valid when the perturbation frequency exceeds 1/2 switching frequency [5]. Sampling and PWM processes will influence the impedance characteristics of the converter in the highfrequency range. Related studies that described the nonlinearity of these processes were mainly developed in the dc-dc system. In [6] and [7], the generated multi-frequency model that considered the PWM sideband harmonic components was used to analyze and construct the high-bandwidth $\mathrm{dc}-\mathrm{dc}$ converters. However, unlike $\mathrm{dc}-\mathrm{dc}$ converters, the ac-dc converters operate at the time-varying operating points. This method cannot be applied directly to ac-dc converters. In [8] and [9], the multi-frequency single-phase and threephase inverter models are obtained to assess the influence of the sampling process. These studies analyzed the effect of generated sampling sideband harmonic components on system stability in a high-frequency range but neglected the PWM sideband harmonic. To accurately evaluate the nonlinear\\[0pt]
characteristics of the PWM process, a two-order admittance model of the three-phase inverter system is derived by considering the PWM sideband harmonic [10]. As demonstrated by the two-order multi-frequency model, the initial phase of the carrier wave was explored under the parallel threephase inverter system. However, the digital PWM model and the effect of semiconductor switching frequency on system stability were not adequately investigated. By combining the method of [8]-[10], a four-order multi-frequency admittance model is constructed in [11] by combining the sampling and PWM processes. The resulting four-order model can be utilized to investigate the HIS problem. However, in [11], the impact of switching frequency on system harmonic stability was not considered. Later, Assefa HY discovered that the switching frequency of semiconductors affects the damping effect in the L-N system [12]. However, the phenomenon was not analyzed in-depth modeling and theoretical study in [12]. The impact of switching frequency on system stability needs to be investigated further. In [13], the switching frequency is taken into account in a matrixbased multi-frequency impedance model of the Buck converter. According to the derived model, it is found that the variation of semiconductor switching frequency could affect the impedance characteristics of the converter and might induce harmonic instability issues. Since the switching frequency of different types of locomotive rectifiers varies from 250 to 1250 Hz , the dampening impact of different switching frequencies on the L-N system should be studied. In conclusion, the present analysis for HIS in the L-N system is still insufficient.

As a result, developing accurate high-frequency locomotive and traction network models that can be used for HIS issues is necessary. In this article, a detailed mathematical model of the PWM process is developed. The dynamic propagations of perturbation frequency and its sideband frequency components in the PWM process are revealed. Then, the locomotive rectifier's multi-frequency input-admittance model is derived. The influence of switching frequency on locomotive impedance characteristics is analyzed. Afterward, based on the multi-frequency model, a modified admittance conversion technique is adopted to transform the multi-frequency model into the single-input-single-output (SISO) model while preserving the sideband frequency couplings. The proposed SISO model is more precise in the frequency range higher than $1 / 2$ switching frequency than the traditional small-signal averaging model. Finally, based on the SISO model, the frequency domain stability criterion is used to assess the stability of the L-N system in various scenarios. The main contributions in this article are summarized as follows.

\begin{enumerate}
  \item The dynamic propagations of the perturbation and its sideband harmonic frequency components in the PWM process are analyzed in detail by the established threeorder PWM transfer function matrix.
  \item The derived SISO locomotive input-admittance model can capture the influence of PWM sideband harmonic coupling and is more accurate above the $1 / 2$ switching frequency.
\end{enumerate}

\begin{figure}[H]
\begin{center}
  \includegraphics[alt={},max width=\columnwidth]{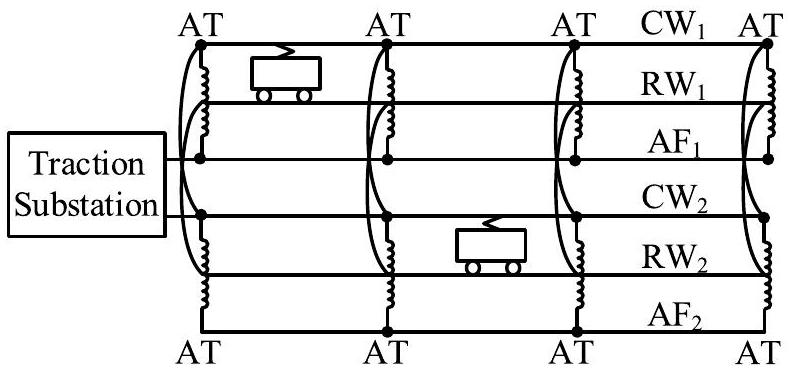}
\caption{Schematic of the all-parallel AT power supply L-N system.}
\end{center}
\end{figure}

\begin{enumerate}
  \setcounter{enumi}{2}
  \item The impact of different semiconductor switching frequencies, control bandwidths, and traction impedance on the L-N system stability is analyzed based on the derived model.\\
The rest of this article is arranged as follows. Section II depicts the researched L-N system and the traction network model. Section III presents the detailed mathematical model of the PWM process and constructs the locomotive inputadmittance model. The stability of the L-N system under different switching frequencies, current control bandwidths, and traction network parameters is discussed in Section IV. The simulation test is then performed on the hardware-in-theloop (HIL) platform. Section V contains the conclusions.
\end{enumerate}

\section*{II. System Description}
At present, the all-parallel autotransformer (AT) traction network is the most common structural form of electrified railways in China [14], [15]. As shown in Fig. 1, the L-N system comprises a traction substation, traction network, AT transformers, locomotives, contact wire (CW), rail wire (RW), and AT feeder (AF), respectively [16]. Uplink and downlink are denoted by subscripts 1 and 2, respectively.

The commonly used traction network models for calculation include the generalized symmetrical components and equivalent circuit method [17]. According to [17], a generalized traction network model of symmetrical components considering AT leakage reactance is shown in Fig. 2. To facilitate analysis, all parameters in Fig. 2 are converted to the low-voltage side of the onboard transformer. $u_{s}$ and $Z_{s}$ are the traction network voltage and traction transformer equivalent impedance. $Z_{\mathrm{AT}}$ is the AT leakage reactance and $Z_{\mathrm{AT}}^{\prime}$ is the equivalent AT leakage reactance in the composite sequence network, equal to $2 Z_{\mathrm{AT}}$; $z_{i}(i=0,1,2,3)$ are the sequence impedance in unit length; $x$ is the distance between the locomotive and the nearest AT station forward; $l$ is the distance between the locomotive to the traction substation; $D_{\mathrm{AT}}$ is the distance of an AT segment.

The equivalent impedance of the all-parallel AT traction network is deduced as

\[
Z_{\mathrm{eq}}=\frac{1}{4 k^{2}}\left[\begin{array}{c}
\left(z_{0} x+Z_{\mathrm{AT}}^{\prime}\right) \frac{z_{0}\left(D_{\mathrm{AT}}-x\right)+Z_{\mathrm{AT}}^{\prime}}{2 Z_{\mathrm{AT}}^{\prime}+z_{0} D_{\mathrm{AT}}}  \tag{1}\\
+z_{1} l+\left(z_{2}+z_{3}\right) x\left(1-\frac{x}{D_{\mathrm{AT}}}\right)
\end{array}\right]
\]

where $k$ is the turns ratio of the onboard transformer.

\begin{figure}[H]
\begin{center}
  \includegraphics[alt={},max width=\columnwidth]{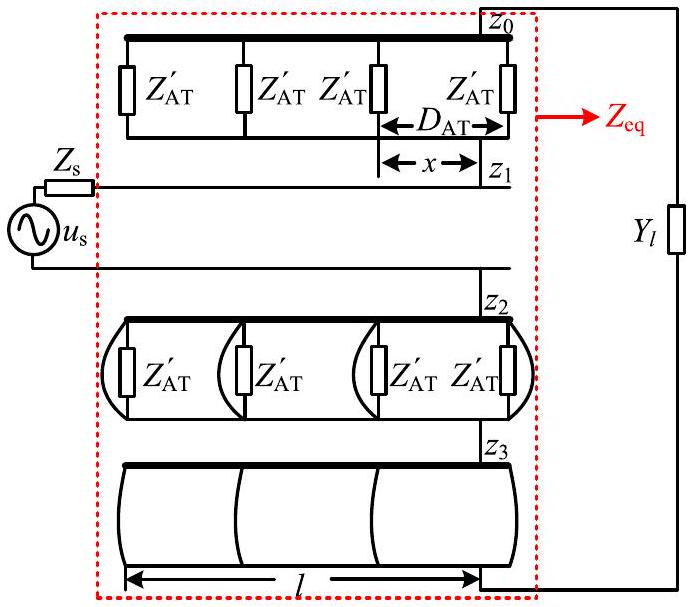}
\caption{Equivalent circuit of the all-parallel AT power supply L-N system.}
\end{center}
\end{figure}

\begin{table}[H]
\begin{center}
\caption{Major Electrical Parameters of the Traction Network}
\begin{tabular}{|l|l|l|}
\hline
Description & Parameters & Values \\
\hline
\multirow{4}{*}{Transformer in traction substation} & Rated ratio & $220 \times(1 \pm 2 \times 2.5 \%)$ \\
\hline
 & Rated power & $50 / 31.5 / 31.5$ MVA \\
\hline
 & Short-circuit voltage & 10.5\% \\
\hline
 & Short-circuit losses & 57.271 kW \\
\hline
\multirow{2}{*}{Traction network} & Equivalent capacitance to ground & $C_{\text{eq }}=2.81 \mathrm{mF}$ \\
\hline
 & Equivalent resistance and inductance & $Z_{\text{eq }}=0.01+\mathrm{j} 1.96 \mathrm{~m} \Omega$ \\
\hline
\multirow{4}{*}{AT Transformer} & Rated ratio & $55 / 27.5 \mathrm{kV}$ \\
\hline
 & Rated power & 32 MVA \\
\hline
 & Short-circuit voltage & 10.5\% \\
\hline
 & Short-circuit losses & 57.271 kW \\
\hline
\end{tabular}
\end{center}
\end{table}

To accurately analyze the stability of the L-N system, the distributed capacitance of the multi-conductor transmission line of the traction network should be considered. In the allparallel AT traction network, the equivalent capacitance of the traction network can be expressed as

\[
\begin{array}{r}
C_{\mathrm{eq}}=k^{2} \sum\left(C_{i \mathrm{CW}}+C_{i \mathrm{AF}}+C_{i \mathrm{RW}}+C_{i \mathrm{PW}}+C_{i \mathrm{GW}}\right) \\
(i=1,2) \tag{2}
\end{array}
\]

where $C_{i \mathrm{CW}}, C_{i \mathrm{AF}}, C_{i \mathrm{RW}}, C_{i \mathrm{PW}}$, and $C_{i \mathrm{GW}}$ denote the capacitance of contact wire, AT feeder wire, rail wire, protect wire, and ground wire to ground, respectively.

The equivalent impedance of the traction transformer is composed of $R_{s}$ and $X_{s}$

\begin{equation*}
R_{s}=P_{k} \times \frac{U_{B}^{2}}{S_{T}^{2}}, \quad X_{s}=\frac{U_{k} \%}{100} \times \frac{U_{B}^{2}}{S_{T}} \tag{3}
\end{equation*}

where $P_{k}, U_{B}, S_{T}, U_{k} \%$ are the short-circuit copper loss, rated voltage, rated capacity, and short-circuit voltage of the transformer.

Since the impedance of the traction network is composed of passive components, the equivalent impedance of the traction network can be calculated as follows based on (1)-(3):

\begin{equation*}
Z_{\mathrm{net}}=\frac{Z_{s}+Z_{\mathrm{eq}}}{s C_{\mathrm{eq}}\left(Z_{s}+Z_{\mathrm{eq}}\right)+1} . \tag{4}
\end{equation*}

The main parameters of the traction power supply system are shown in Table I [14].

\section*{III. Model of Locomotive Input-Admittance}
As shown in Fig. 3, the complete control in the locomotive rectifier includes the phase-locked loop (PLL), the direct voltage control (DVC), the alternating current control (ACC), and the PWM block. Since the PLL and the DVC have a relatively low control bandwidth, these control blocks significantly impact the locomotive input-impedance in the low-frequency range [5]. The modulation time delay and the generated PWM sideband harmonics in the PWM block significantly impact the locomotive input-admittance in the high-frequency range. In this article, the main research object

\begin{figure}[H]
\begin{center}
  \includegraphics[alt={},max width=\columnwidth]{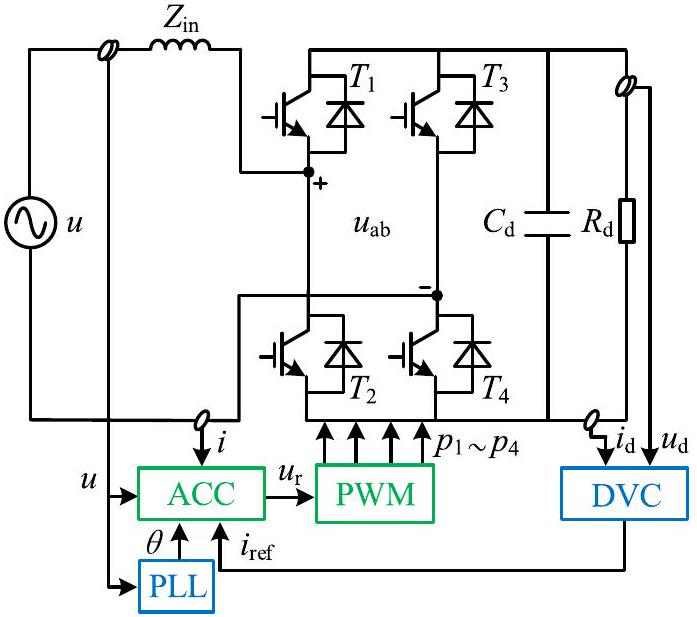}
\caption{Control block diagram of the locomotive rectifier.}
\end{center}
\end{figure}

\begin{figure}[H]
\begin{center}
  \includegraphics[alt={},max width=\columnwidth]{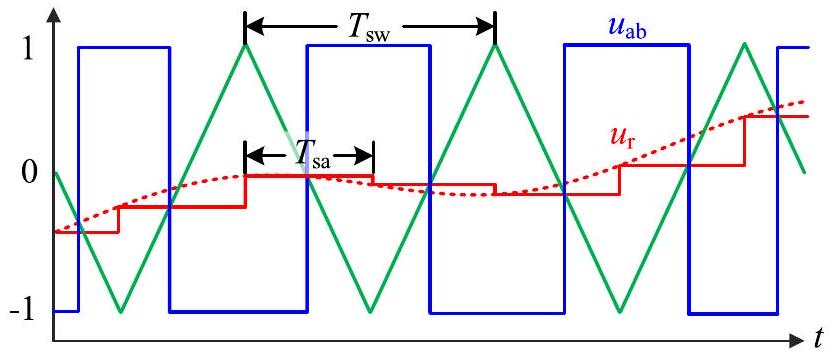}
\caption{Schematic of digital PWM process.}
\end{center}
\end{figure}

is to analyze the harmonic instability issues in the $\mathrm{L}-\mathrm{N}$ system. When the harmonic instability issues occur, the main harmonic components are commonly five times higher than the fundamental frequency [3]. Therefore, in order to simplify the input-admittance model of locomotive, the control blocks that have a slight impact on the high-frequency range, such as PLL and DVC, can be omitted.

\section*{A. Model of Digital PWM Comparator}
Fig. 4 depicts a schematic of the digital PWM process using a bipolar asymmetric regular sampling method with the

\begin{figure}[H]
\begin{center}
  \includegraphics[alt={},max width=\columnwidth]{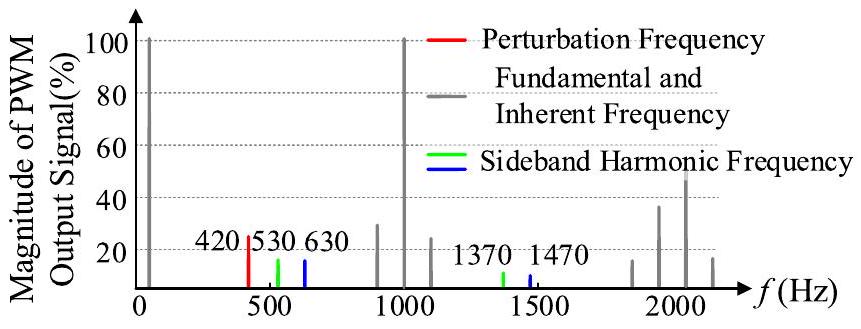}
\caption{Frequency spectrum of the PWM output signal when PWM reference signal contains perturbation frequency at 420 Hz .}
\end{center}
\end{figure}

sampling occurring at the triangle carrier's peak [18], i.e., $f_{\mathrm{sa}}=2 f_{\mathrm{sw}}$. The green line represents the carrier wave, and the blue line represents the PWM output signal, where $T_{\mathrm{sw}}$ is the switching period. The continuous PWM reference signal and the discrete PWM reference signal following the sample and hold (S\&H) process are shown by the red dotted line and the solid red line, respectively.

The perturbation frequency component of the PWM reference signal in the time domain results in the double convolution of sideband harmonics in the frequency domain, and endless sideband harmonic components are generated in the PWM output signal due to the nonlinearity of the PWM process [19]. As shown in Fig. 5, when the perturbation frequency components in the PWM reference signal are at 420 Hz , the harmonic components in the PWM output signal are perturbation frequency ( 420 Hz ), fundamental frequency $(50 \mathrm{~Hz})$, inherent switching frequency harmonics ( 900,1000 , 1100 Hz , etc.), and sideband harmonic frequency components ( $530,630,1370$, and 1470 Hz ). The sideband harmonic frequency can be summarized as follows:

\[
\begin{array}{ll}
f_{\mathrm{pwmb} 1}=f_{\mathrm{sw}}-f_{p}-f_{0}, & f_{\mathrm{pwmb} 2}=f_{\mathrm{sw}}-f_{p}+f_{0} \\
f_{\mathrm{pwmb} 3}=f_{\mathrm{sw}}+f_{p}-f_{0}, & f_{\mathrm{pwmb} 4}=f_{\mathrm{sw}}+f_{p}+f_{0} \tag{6}
\end{array}
\]

According to the Nyquist-Shannon sampling theorem, the aliasing phenomenon will occur when the perturbation frequency exceeds the Nyquist frequency [18]. To avoid the aliasing phenomenon, the analyzed upper-frequency range in this article is the Nyquist frequency. As a result, the sideband harmonic components of $f_{\mathrm{pwmb} 1}, f_{\mathrm{pwmb} 2}$ are only considered in this study, but $f_{\mathrm{pwmb3}}$ and $f_{\mathrm{pwmb} 4}$ are ignored.

Assuming the initial phase angle of the PWM carrier waveform is set to $0^{\circ}$, the PWM reference signal can be expressed as

\begin{equation*}
u_{r}=M_{0} \cos \left(\omega_{0} t+\theta_{0}\right)+M_{p} \cos \left(\omega_{p} t+\theta_{p}\right) \tag{7}
\end{equation*}

where the variable symbols $x_{0}$ and $x_{p}$ denote the fundamental and perturbation frequency components, respectively.

As derived in (8), as shown at the bottom of the page, the mathematical expression of the PWM output signal in the frequency domain is generated using 1-D spectrum analysis [19], where $\alpha(\omega)=2 \omega_{\mathrm{sw}} / \pi \omega, \beta(\omega)=\pi / 2 m, \gamma(\omega)=\pi \omega / 2 \omega_{\mathrm{sw}}$. $J_{n}$ is the first kind of Bessel function in the order $n$, which can be expressed as follows:

\begin{equation*}
J_{n}(x)=\frac{1}{2 \pi} \int_{0}^{2 \pi} \cos (n \tau-x \sin \tau) d \tau \tag{9}
\end{equation*}

The harmonic balance principle can be used to derive the relationship between the perturbation frequency and its sideband harmonic frequency components [20], as shown in (10) and (11). $G_{0}$ can be obtained by setting $m, n$, and $k$ to 0,1 , and 1 , respectively. $G_{0}$ is the mapping from the PWM reference signal's perturbation frequency to the same frequency in the PWM output signal

\begin{equation*}
G_{0}(s)=J_{0}\left(\frac{\pi}{2} \frac{|s|}{\omega_{\mathrm{sw}}} M_{0}\right) . \tag{10}
\end{equation*}

$G_{1}$ and $G_{2}$ can be obtained similarly when $m, n$, and $k$ are equal to $1, \pm 1$, and -1 , respectively. $G_{1}$ and $G_{2}$, which describe the dynamic frequency coupling in the PWM process, are mappings from the perturbation frequency in $U_{r}(\omega)$ to its sideband frequency components in $P(\omega)$

\begin{align*}
& G_{1}(s)=-J_{1}\left(\frac{\pi}{2} M_{0} \frac{\omega_{\mathrm{sw}}-|s|-\omega_{0}}{\omega_{\mathrm{sw}}}\right) e^{-j \theta_{0}} \\
& G_{2}(s)=-J_{1}\left(\frac{\pi}{2} M_{0} \frac{\omega_{\mathrm{sw}}-|s|+\omega_{0}}{\omega_{\mathrm{sw}}}\right) e^{j \theta_{0}} \tag{11}
\end{align*}

The dynamic propagations of these frequency components in the PWM process are shown in Fig. 6.

According to Fig. 6, the PWM model can be established as follows: $G_{d}(s)=\exp \left(-0.5 T_{\mathrm{sa}}-\lambda T_{\mathrm{sa}}\right) s, T_{\mathrm{sa}}$ is the sampling time delay, $\lambda$ is the computation time delay in the digital process

\begin{align*}
{\left[\begin{array}{c}
P\left(f_{p}\right) \\
P\left(f_{\mathrm{pwmb} 1}\right) \\
P\left(f_{\mathrm{pwmb} 2}\right)
\end{array}\right] } & =\vec{G}_{d} \vec{G}_{\mathrm{pwm}}\left[\begin{array}{c}
U_{r}\left(f_{p}\right) \\
U_{r}\left(f_{\mathrm{pwmb} 1}\right) \\
U_{r}\left(f_{\mathrm{pwmb} 2}\right)
\end{array}\right]  \tag{12}\\
\vec{G}_{\mathrm{pwm}} & =\left[\begin{array}{ccc}
G_{0}\left(f_{p}\right) & G_{1}\left(f_{\mathrm{pwmb} 1}\right) & G_{2}\left(f_{\mathrm{pwmb} 2}\right) \\
G_{1}\left(f_{p}\right) & G_{0}\left(f_{\mathrm{pwmb} 1}\right) & 0 \\
G_{2}\left(f_{p}\right) & 0 & G_{0}\left(f_{\mathrm{pwmb} 2}\right)
\end{array}\right]  \tag{13}\\
\vec{G}_{d} & =\left[\begin{array}{ccc}
G_{d}\left(f_{p}\right) & 0 & 0 \\
0 & G_{d}\left(f_{\mathrm{pwmb} 1}\right) & 0 \\
0 & 0 & G_{d}\left(f_{\mathrm{pwmb} 2}\right)
\end{array}\right] . \tag{14}
\end{align*}

The modeled and measured responses of $G_{0}, G_{1}$, and $G_{2}$ at frequency up to $2 f_{\text{sw }}$ are scanned in Simulink to validate the

\begin{align*}
P(\omega)= & \sum_{m=-\infty}^{\infty} \sum_{n=-\infty}^{\infty} \sum_{\substack{k=-\infty \\
k \in n_{p} Z}}^{\infty} \frac{\pi}{j}\binom{J_{k / n_{p}}\left(\gamma\left(m \omega_{\mathrm{sw}}+n \omega_{0}\right) M_{p}\right)}{\cdot\binom{j^{n-k+k / n_{p}} J_{n-k}\left(\gamma\left(m \omega_{\mathrm{sw}}+n \omega_{0}\right) M_{0}\right) e^{j \beta\left(m \omega_{\mathrm{sw}}+n \omega_{0}\right)}}{-j^{-n-k+k / n_{p}} J_{n+k}\left(\gamma\left(m \omega_{\mathrm{sw}}+n \omega_{0}\right) M_{0}\right) e^{-j \beta\left(m \omega_{\mathrm{sw}}+n \omega_{0}\right)}}} \\
& \cdot \alpha\left(m \omega_{\mathrm{sw}}+n \omega_{0}\right) \cdot \delta\left(\omega-m \omega_{\mathrm{sw}}-n \omega_{0}\right) \tag{8}
\end{align*}

\begin{figure}[H]
\begin{center}
  \includegraphics[alt={},max width=\columnwidth]{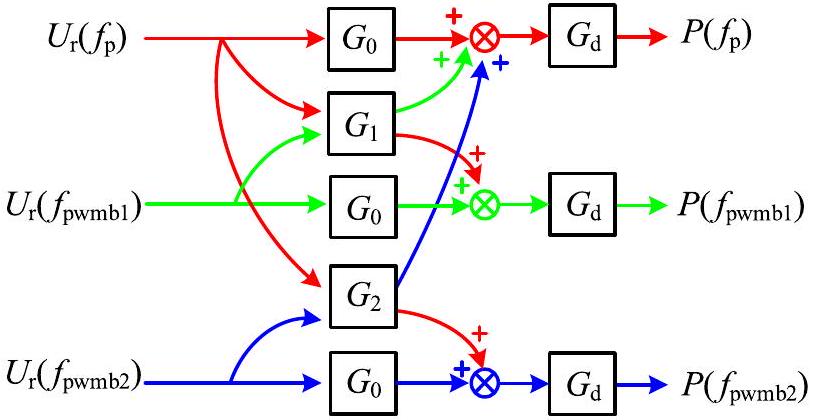}
\caption{Frequency-domain dynamic propagations in the PWM process.}
\end{center}
\end{figure}

\begin{figure}[H]
\begin{center}
  \includegraphics[alt={},max width=\columnwidth]{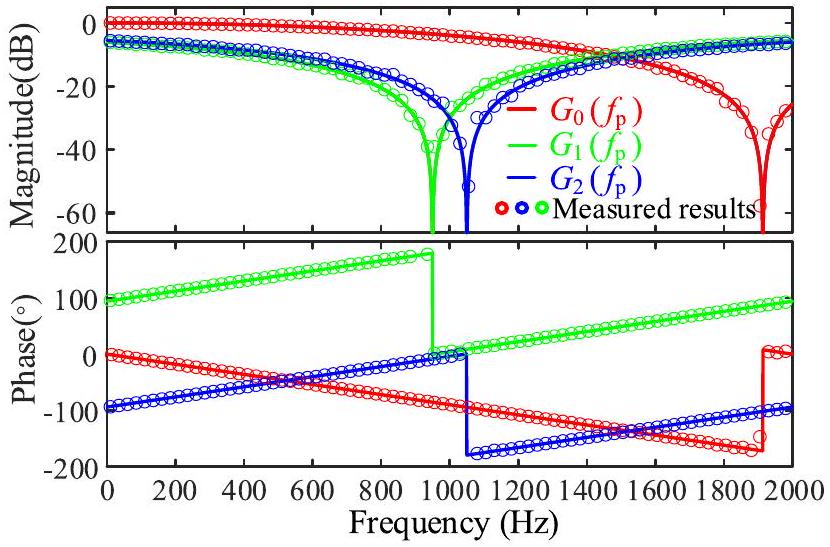}
\caption{Modeled and measured results of $G_{0}, G_{1}, G_{2}$.}
\end{center}
\end{figure}

transfer function analytical results in (10) and (11). As shown in Fig. 7, the red, green, and blue lines denote $G_{0}, G_{1}$, and $G_{2}$, respectively. The solid lines reflect the modeled results, and the circles represent the measured results. The results reveal that the measured results agree well with the theoretical model.

\section*{B. Multi-Frequency Input-Admittance Model of Locomotive Rectifier}
The equivalent admittance of the onboard transformer ( $Y_{\text{in }}$ ), the ACC proportional integral (PI) regulator ( $G_{\text{acc }}$ ), are linear elements. As a result, they can be expressed as a diagonal matrix

\begin{align*}
\vec{Y}_{\mathrm{in}} & =\left[\begin{array}{ccc}
Y_{\mathrm{in}}\left(f_{p}\right) & 0 & 0 \\
0 & Y_{\mathrm{in}}\left(f_{\mathrm{pwmb} 1}\right) & 0 \\
0 & 0 & Y_{\mathrm{in}}\left(f_{\mathrm{pwmb} 2}\right)
\end{array}\right]  \tag{15}\\
\vec{G}_{\mathrm{acc}} & =\left[\begin{array}{ccc}
G_{\mathrm{acc}}\left(f_{p}\right) & 0 & 0 \\
0 & G_{\mathrm{acc}}\left(f_{\mathrm{pwmb} 1}\right) & 0 \\
0 & 0 & G_{\mathrm{acc}}\left(f_{\mathrm{pwmb} 2}\right)
\end{array}\right] . \tag{16}
\end{align*}

Based on Fig. 8, the voltage and current can be given as follows:

\begin{align*}
\vec{I}(s) & =\vec{Y}_{\mathrm{in}}(s)\left[\vec{U}(s)-\vec{U}_{\mathrm{ab}}(s)\right]  \tag{17}\\
\vec{U}_{\mathrm{ab}}(s) & =\vec{G}_{\mathrm{pwm}}(s) \vec{U}(s)+\vec{G}_{\mathrm{pwm}}(s) \vec{G}_{\mathrm{acc}}(s) \vec{I}(s) \tag{18}
\end{align*}

Then, according to (12)-(18), the multi-frequency input-admittance model of locomotive rectifier is derived

\begin{figure}[H]
\begin{center}
  \includegraphics[alt={},max width=\columnwidth]{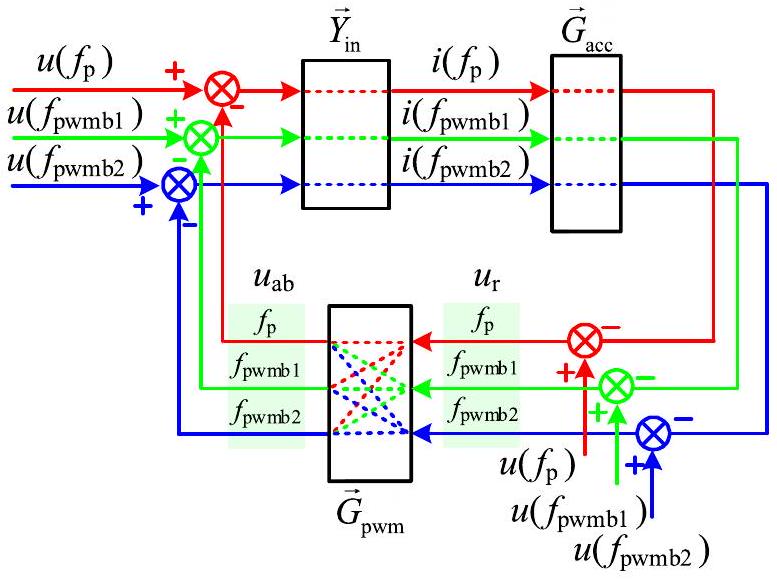}
\caption{Block diagram for the locomotive rectifier closed-loop control system.}
\end{center}
\end{figure}

as follows:

\begin{align*}
\vec{Y}_{\mathrm{rec}}(s) & =\vec{I}(s) / \vec{U}(s) \\
& =\frac{\vec{Y}_{\mathrm{in}}(s)-\vec{Y}_{\mathrm{in}}(s) \vec{G}_{\mathrm{pwm}}(s)}{\vec{E}(s)+\vec{Y}_{\mathrm{in}}(s) \vec{G}_{\mathrm{acc}}(s) \vec{G}_{\mathrm{pwm}}(s)} \tag{19}
\end{align*}

where $\vec{E}(s)$ is a three-order unit matrix. The input-admittance of the locomotive rectifier is a multi-frequency matrix and can be expressed in the following equation when the perturbation frequency $f_{p}$, sideband harmonics $f_{\text{pwmb1 }}$, and $f_{\text{pwmb2 }}$ are considered:

\[
\vec{Y}_{\mathrm{rec}}(s)=\left[\begin{array}{ccc}
Y_{\mathrm{rec}, m}\left(f_{p}\right) & Y_{\mathrm{rec}, b 1}\left(f_{\mathrm{pwmb} 1}\right) & Y_{\mathrm{rec}, b 2}\left(f_{\mathrm{pwmb} 2}\right)  \tag{20}\\
Y_{\mathrm{rec}, b 1}\left(f_{p}\right) & Y_{\mathrm{rec}, m}\left(f_{\mathrm{pwmb} 1}\right) & 0 \\
Y_{\mathrm{rec}, b 2}\left(f_{p}\right) & 0 & Y_{\mathrm{rec}, m}\left(f_{\mathrm{pwmb} 2}\right)
\end{array}\right] .
\]

When the perturbation voltages $U\left(f_{p}\right)$ are only injected, the transfer functions of $Y_{\text{rec }, m}\left(f_{p}\right)$ can be further given as follows:

\begin{align*}
Y_{\mathrm{rec}, m}\left(f_{p}\right) & \frac{Y_{\mathrm{in}}\left(f_{p}\right)-G_{0}\left(f_{p}\right) G_{d}\left(f_{p}\right) Y_{\mathrm{in}}\left(f_{p}\right)}{1+T_{\mathrm{op}}\left(f_{p}\right)} \\
& -\frac{G_{1}\left(f_{\mathrm{pwmb} 1}\right) G_{d}\left(f_{\mathrm{pwmb} 1}\right) G_{\mathrm{acc}}\left(f_{p}\right) Y_{\mathrm{in}}\left(f_{p}\right)}{1+T_{\mathrm{op}}\left(f_{p}\right)} Y_{\mathrm{rec}, b 1}\left(f_{p}\right) \\
& -\frac{G_{2}\left(f_{\mathrm{pwmb} 2}\right) G_{d}\left(f_{\mathrm{pwmb} 2}\right) G_{\mathrm{acc}}\left(f_{p}\right) Y_{\mathrm{in}}\left(f_{p}\right)}{1+T_{\mathrm{op}}\left(f_{p}\right)} Y_{\mathrm{rec}, b 2}\left(f_{p}\right)
\end{align*}

where $T_{\mathrm{op}}\left(f_{p}\right)=G_{0}\left(f_{p}\right) G_{d}\left(f_{p}\right) G_{\mathrm{acc}}\left(f_{p}\right) Y_{\mathrm{in}}\left(f_{p}\right)$. The first item in (21) is the small-signal averaging model. The second and the third items in (21) reflect the influence of the PWM sideband harmonics. It is shown that the sideband harmonic terms can be regarded as an additional parallel admittance added to the small-signal averaging impedance model.

To verify the correctness of the theoretical model results, the admittance measurement circuit needs to be established. The main parameters of the locomotive are shown in Table II. As shown in Fig. 9, the frequency at 560 Hz (approximately equal to $1 / 2$ switching frequency) is a separatrix. The magnitude of $Y_{\text{rec }, m}\left(f_{p}\right)$ is greater than $Y_{\text{rec }, b 1}\left(f_{p}\right)$ and $Y_{\text{rec }, b 2}\left(f_{p}\right)$

\begin{table}[H]
\begin{center}
\caption{Major Electrical Parameters of the Locomotive}
\begin{tabular}{|l|l|l|}
\hline
Symbol & Description & Value \\
\hline
$P_{\mathrm{n}}$ & Rated power of one power unit & 340 kW \\
\hline
$U_{\mathrm{n}}$ & Rated voltage (low-voltage side of the onboard transformer) & 1500 V \\
\hline
$f_{0}$ & Rated fundamental frequency & 50 Hz \\
\hline
$L_{\text{in }}$ & Equivalent impedance of onboard transformer & 2.1 mH \\
\hline
$K_{\mathrm{ip}} / K_{\mathrm{ii}}$ & Gain of ACC PI regulator & $0.76 / 10 \Omega$ \\
\hline
$f_{\mathrm{sw}}$ & Switching frequency & 1 kHz \\
\hline
$f_{\mathrm{sa}}$ & Sampling frequency & 2 kHz \\
\hline
\end{tabular}
\end{center}
\end{table}

\begin{figure}[H]
\begin{center}
  \includegraphics[alt={},max width=\columnwidth]{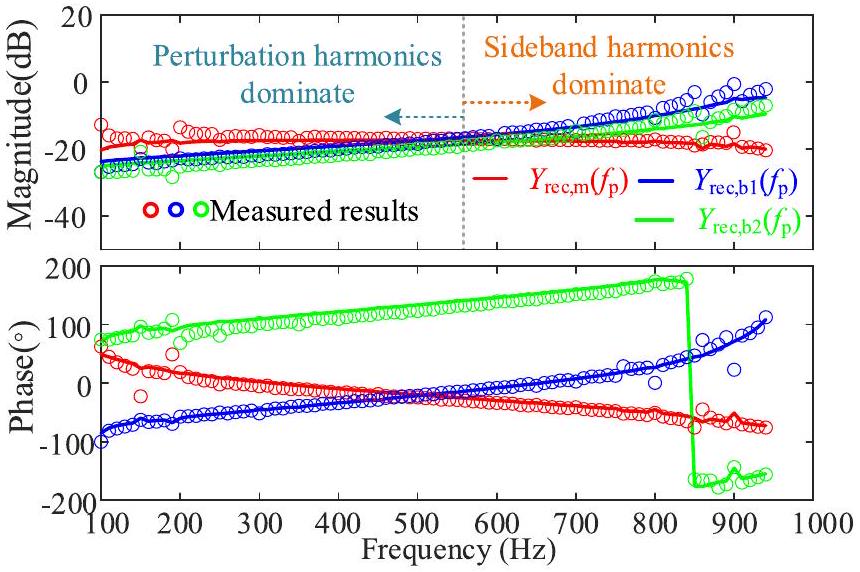}
\caption{Derived multi-frequency model and the admittance measurement results of locomotive rectifier.}
\end{center}
\end{figure}

when the perturbation frequency is lower than the separatrix, indicating that the perturbation harmonic component dominates the locomotive rectifier input-admittance. However, the magnitude of $Y_{\text{rec }, b 1}\left(f_{p}\right)$ and $Y_{\text{rec }, b 2}\left(f_{p}\right)$ will be greater than $Y_{\text{rec }, m}\left(f_{p}\right)$ when the perturbation frequency is higher than the separatrix, indicating that sideband harmonic components dominate the locomotive rectifier input-admittance. The farther the perturbation frequency is from the separatrix, the more significant the magnitude gap between the perturbation harmonic and its sideband components.

It is worth mentioning that the traditional small-signal averaging model overlooks the impact of PWM sideband harmonic components. As a result, the typical converter model's effective frequency range is less than $1 / 2$ switching frequency. The PWM sideband harmonic components should be considered when the locomotive rectifier's high-frequency input-admittance characteristic is analyzed. The separatrix could be explained as follows. According to (5), when the perturbation frequency is higher than the separatrix, the generated PWM sideband frequency components are lower than the perturbation frequency. A low-pass filter is provided by the closed-loop control system, which includes the onboard transformer equivalent impedance and the ACC regulator. As a result, the magnitude of sideband frequency components is less attenuated than the perturbation harmonic when the perturbation frequency exceeds the separatrix.

\begin{figure}[H]
\begin{center}
  \includegraphics[alt={},max width=\columnwidth]{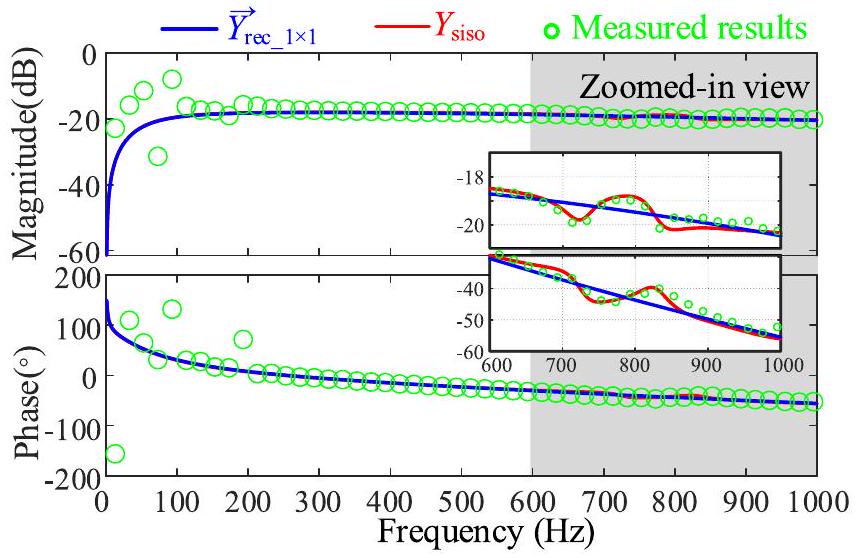}
\caption{Comparison of the traditional model, the proposed SISO model, and the measured admittance results of the locomotive rectifier.}
\end{center}
\end{figure}

\section*{C. Equivalent SISO Input-Admittance Model of Locomotive Rectifier}
Due to the presence of multi-eigenloci, frequency-domain stability analysis methods such as the generalized Nyquist criterion are not intuitive under multi-input-multi-output (MIMO) impedance model [21]. The derived MIMO model can be transformed into a SISO model and simplify the stability analysis results while keeping the influence of PWM sideband harmonic coupling. The technique that converts an MIMO model into a SISO model [21] is adopted in this article

\begin{equation*}
Y_{\text{siso }}(s)=\vec{Y}_{\text{rec\_ } 1 \times 1}-\alpha_{1 \times 2}\left(\vec{E}+c_{2 \times 2} Q_{2 \times 2}\right)^{-1} c_{2 \times 2} b_{2 \times 1} \tag{22}
\end{equation*}

where $\alpha, c, b$ can be obtained from $\vec{Y}_{\text{rec }}$ in (20), $Q$ can be obtained from $\vec{Z}_{\text{net }}$, and the subscript $m \times n$ represents the length of row and column for corresponding element, $\vec{E}$ is a two-order unit matrix.

As shown in Fig. 10, the small-signal averaging model, the SISO model, and the admittance measurement results are compared. It is found that the SISO model, especially in the high-frequency region ( $700-1000 \mathrm{~Hz}$ ), fits measured results better than the small-signal averaging model. Since the PLL and voltage outer loop are disregarded while modeling, the accuracy of the SISO model in the low-frequency range $(0-100 \mathrm{~Hz})$ is low. It can be observed that the derived SISO model is more accurate in the mid-frequency and highfrequency ranges, which can be used to analyze the harmonic instability issues in the L-N system.

\section*{D. Input-Admittance Model of Locomotive}
As shown in Fig. 11, one locomotive has four identical power units. The onboard transformer first steps down the voltage in the traction network from 27.5 kV to 950 V , and then the 950 V ac voltage is converted into 1800 V dc voltage through the locomotive rectifier. The dc voltage remains constant at 1800 V because of the dc-side filter capacitance. Then, considering that the locomotive operates in constant power, the inverter and the traction motor could be equivalent to a constant load resistance. Assuming the

\begin{figure}[H]
\begin{center}
  \includegraphics[alt={},max width=\columnwidth]{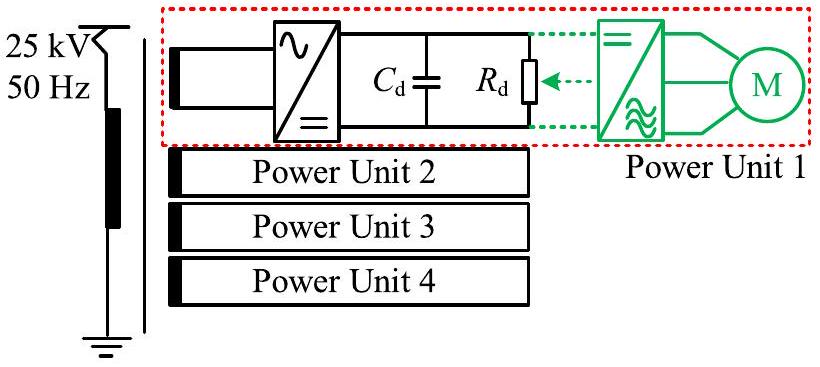}
\caption{Schematic of traction power supply units in locomotive.}
\end{center}
\end{figure}

\begin{figure}[H]
\begin{center}
  \includegraphics[alt={},max width=\columnwidth]{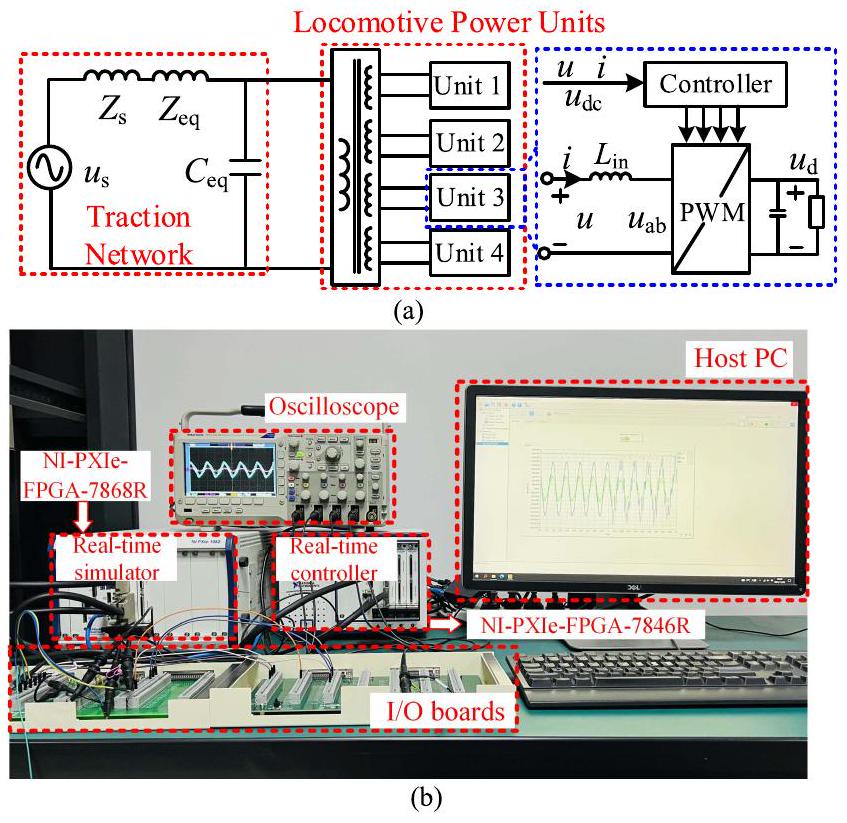}
\caption{HIL simulation platform setup. (a) Circuit model of the L-N system. (b) Configuration of the HIL platform.}
\end{center}
\end{figure}

four power units are connected in parallel and operated independently, the input-impedance of one locomotive can be expressed as $Y_{l}=4 Y_{\text{siso }}$.

\section*{IV. HIL Simulation and Verification}
To verify the derived model and stability analysis results, the HIL simulation platform is built. As shown in Fig. 12(a), the locomotive is represented by four power units with identical structures, and the traction network is represented by the red dotted frame. As shown in Fig. 12(b), the HIL platform is composed of a real-time simulator (NI-PXIe-FPGA-7868R), a host PC, an oscilloscope, a real-time controller (NI-PXIe-FPGA-7846R), and I/O boards. The power units of the locomotive and traction network are operated in the real-time simulator with $1 \mu \mathrm{~s}$ time step. The control strategy of the locomotive power unit is performed in the real-time controller. The real-time simulator converts the simulated analog signal into a range of $-10-10 \mathrm{~V}$ and transmits it into the I/O boards. The analog signals, such as voltage and current at the point of common coupling (PCC), are sampled by the internal analog to digital converter (ADC) module in the real-time controller with the sampling frequency $f_{\mathrm{sa}}=2 f_{\mathrm{sw}}$. Then, the calculated PWM pulses would be\\
sent to the real-time simulator for controlling the locomotive rectifier. Besides, the voltage and current waveforms at PCC are displayed and analyzed on the oscilloscope. The entire system can be controlled by the software StarSim on the host PC.

\section*{A. Influence of Switching Frequency on System Stability}
Based on the established impedance model of traction network and locomotive, the impedance ratio of the $\mathrm{L}-\mathrm{N}$ system can be formulated as follows, where $k$ is the ratio of the onboard transformer:

\begin{equation*}
L(s)=Z_{\text{net }} Y_{l}=4 Z_{\text{net }} Y_{\text{siso }} . \tag{23}
\end{equation*}

Assuming that the locomotive and traction network are in a stable state under independent operation, there are no right-half-plane poles in $Y_{l}$ and $Z_{\text{net }}$. In Fig. 13, the Nyquist diagrams are drawn to intuitively compare the influence of different switching frequencies on system stability. In Fig. 13(a), when $f_{\mathrm{sw}}$ is set to 1000 Hz , the Nyquist diagram is far from the critical point ( $-1, j 0$ ), and the stability margin is wide enough. Then, as shown in Fig. 13(b) and (c), when $f_{\text{sw }}$ is decreased to 800 and 600 Hz , the Nyquist diagram gradually close to the critical point $(-1, j 0)$. It can be observed that the system is in a critical stable state when $f_{\text{sw }}$ is set to 600 Hz . Moreover, as shown in Fig. 13(d), when $f_{\mathrm{sw}}$ is set to 500 Hz , the Nyquist diagram encircles the critical point, indicating that the system is unstable.

The HIL results under various switching frequencies are shown in Fig. 14. In Fig. 14(a), the harmonic components are low, and the system is basically stable. Then, as shown in Fig. 14(b), the waveforms are distorted when the switching frequency is changed from 1000 to 600 Hz . This result supports the conclusion that the system is critically stable when $f_{\mathrm{sw}}$ is set to 600 Hz . Furthermore, when the switching frequency is decreased from 1000 to 500 Hz in Fig. 14(c), the harmonic components in voltage and current will be considerably amplified. The system is unstable due to the unstable PCC voltage. In conclusion, the HIL results match the stability analysis results well. The derived SISO model of the locomotive can analyze the impact of the switching frequency on system stability margin.

The effect of switching frequency on the system stability can be summarized as follows. First, the sampling instants are determined by the switching frequency under the regular sampling strategy [18]. As the switching frequency decreases, the sampling period of the controller increases, resulting in a larger sampling control time delay. This might lead to a more negative damping range of the locomotive input-admittance in the high-frequency range. Therefore, the system is more prone to the high-frequency HIS. Second, the distribution of PWM sideband harmonic components will be changed with the switching frequency as shown in (5). The onboard locomotive transformer and ACC PI regulator act as the low-pass filter and have a weak attenuation effect on lowfrequency harmonics. Therefore, the generated PWM sideband harmonics will interfere with the control performance under a lower switching frequency. As a result, more harmonics will

\begin{figure}[H]
\begin{center}
  \includegraphics[alt={},max width=\columnwidth]{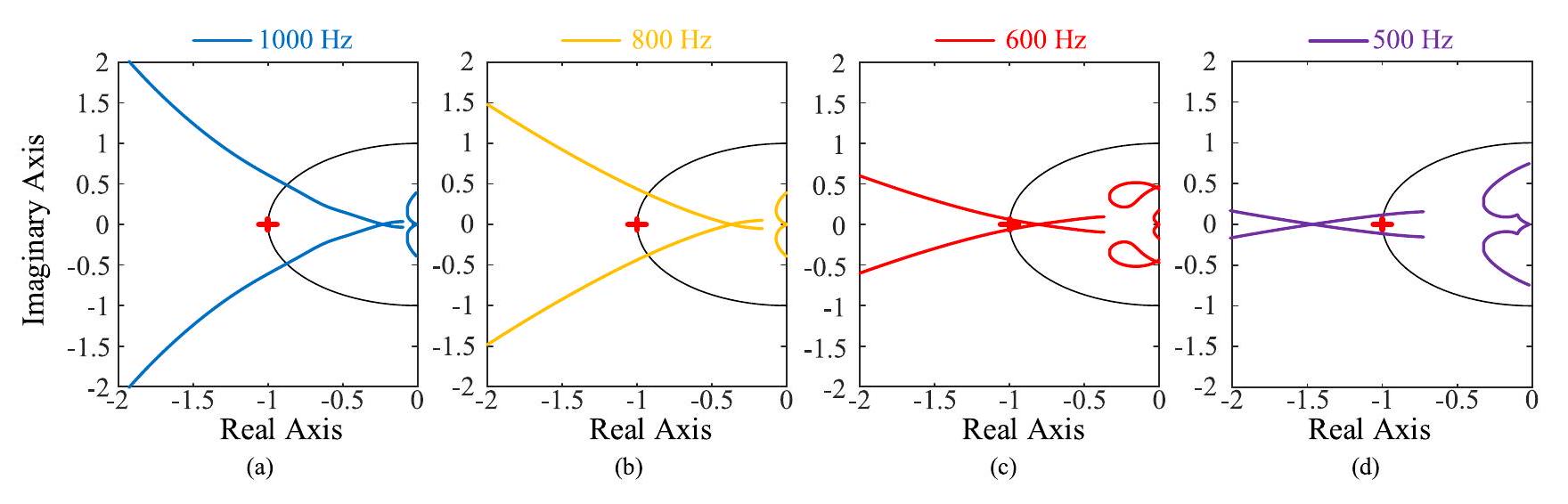}
\caption{Nyquist diagram of impedance ratio under different switching frequencies. (a) $f_{\mathrm{sw}}=1000 \mathrm{~Hz}$. (b) $f_{\mathrm{sw}}=800 \mathrm{~Hz}$. (c) $f_{\mathrm{sw}}=600 \mathrm{~Hz}$. (d) $f_{\mathrm{sw}}=500 \mathrm{~Hz}$.}
\end{center}
\end{figure}

\begin{figure}[H]
\begin{center}
  \includegraphics[alt={},max width=\columnwidth]{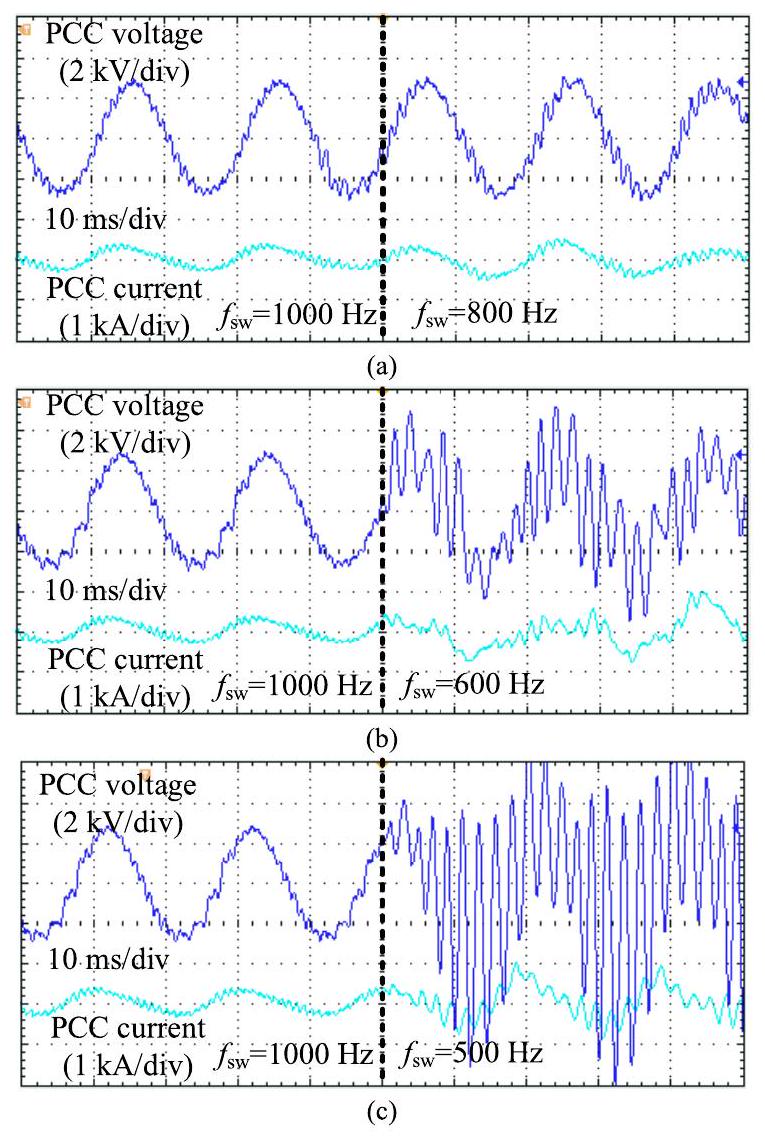}
\caption{Voltage and current waveforms under different switching frequencies. (a) $f_{\text{sw }}$ decreases from 1000 to 800 Hz . (b) $f_{\text{sw }}$ decreases from 1000 to 600 Hz . (c) $f_{\text{sw }}$ decreases from 1000 to 500 Hz .}
\end{center}
\end{figure}

be contained in the ac waveforms. Thus, in order to weaken the influence of PWM sideband harmonics, the control bandwidth should be designed to be less than $10 \%$ of the switching frequency to provide better control performance [19].

\section*{B. Influence of Control Bandwidth and Traction Network Impedance on System Stability}
According to the established model in (4) and (21), the effect of current control bandwidth and traction impedance

\begin{figure}[H]
\begin{center}
  \includegraphics[alt={},max width=\columnwidth]{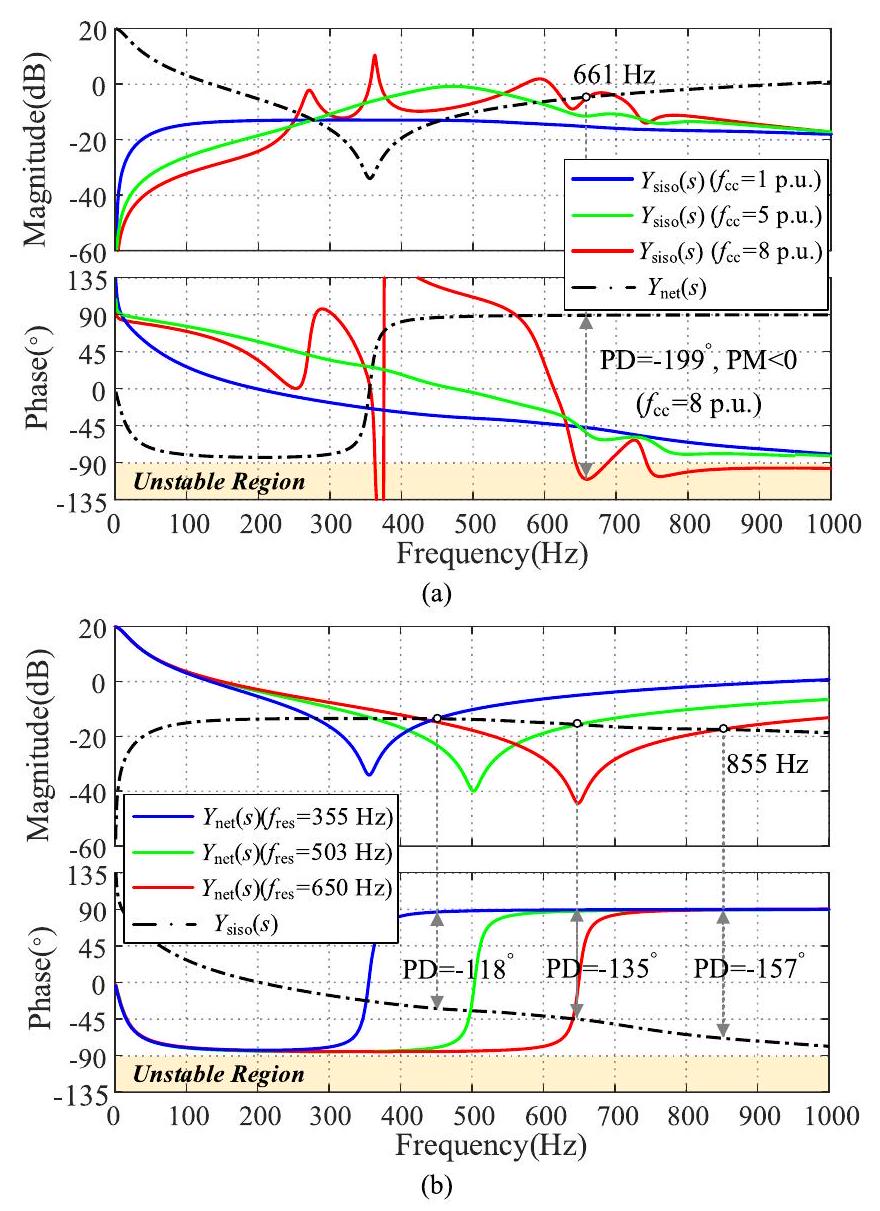}
\caption{Bode diagram of $Y_{\text{siso }}(s)$ and $Y_{\text{net }}(s)$. (a) Influence of different current control bandwidths. (b) Influence of different traction impedance.}
\end{center}
\end{figure}

on the stability boundary is shown in Fig. 15. The intersection points of the traction network and locomotive input-admittance can be used to evaluate system stability. The L-N system will be stable when the phase difference (PD) is less than $180^{\circ}$, and the phase margin (PM) is larger than $0^{\circ}$ at the intersection points. To analysis the stability results more clearly, the intersection points of locomotive and traction network admittance under different parameters are summarized in Table III.

\begin{table}[H]
\begin{center}
\caption{System Stability Analysis Under Different Parameters}
\begin{tabular}{|l|l|l|l|}
\hline
Types & Parameters & Frequency of intersection points (PD \& PM) & Mode \\
\hline
\multirow{3}{*}{ACC Control Bandwidth} & $f_{\mathrm{cc}}=1$ p.u. & $281 \mathrm{~Hz}\left(68.2^{\circ} \& 111.8^{\circ}\right) ; 449 \mathrm{~Hz}\left(-118^{\circ} \& 62^{\circ}\right)$ & Stable \\
\hline
 & $f_{\mathrm{cc}}=5$ p.u. & $278 \mathrm{~Hz}\left(116.3^{\circ}\right.$ \& $\left.63.7^{\circ}\right) ; 585 \mathrm{~Hz}\left(-116.7^{\circ}\right.$ \& $\left.63.3^{\circ}\right)$ & Stable \\
\hline
 & $f_{\mathrm{cc}}=8$ p.u. & $255 \mathrm{~Hz}\left(83.4^{\circ} \& 96.6^{\circ}\right) ; 661 \mathrm{~Hz}\left(-199.2^{\circ} \&-19.2^{\circ}\right)$; \( 695 \mathrm{~Hz}\left(-173.4^{\circ} \& 6.6^{\circ}\right) \) & Unstable \\
\hline
\multirow{3}{*}{Traction Network Parameter (Resonance frequency)} & $0.1 \Omega / 1 \mathrm{mH} / 200 \mu \mathrm{~F}(432 \mathrm{~Hz})$ & $281 \mathrm{~Hz}\left(68.2^{\circ}\right.$ \& $\left.111.8^{\circ}\right) ; 449 \mathrm{~Hz}\left(-118^{\circ}\right.$ \& $\left.62^{\circ}\right)$ & Stable \\
\hline
 & $0.1 \Omega / 1 \mathrm{mH} / 100 \mu \mathrm{~F}(611 \mathrm{~Hz})$ & $352 \mathrm{~Hz}\left(70.8^{\circ} \& 109.2^{\circ}\right) ; 643 \mathrm{~Hz}\left(-135.7^{\circ} \& 44.3^{\circ}\right)$ & Stable \\
\hline
 & $0.1 \Omega / 1 \mathrm{mH} / 60 \mu \mathrm{~F}(789 \mathrm{~Hz})$ & $454 \mathrm{~Hz}\left(59.8^{\circ}\right.$ \& $\left.120.2^{\circ}\right) ; 855 \mathrm{~Hz}\left(-157.7^{\circ}\right.$ \& $\left.22.3^{\circ}\right)$ & Critical stable \\
\hline
\end{tabular}
\end{center}
\end{table}

\begin{figure}[H]
\begin{center}
  \includegraphics[alt={},max width=\columnwidth]{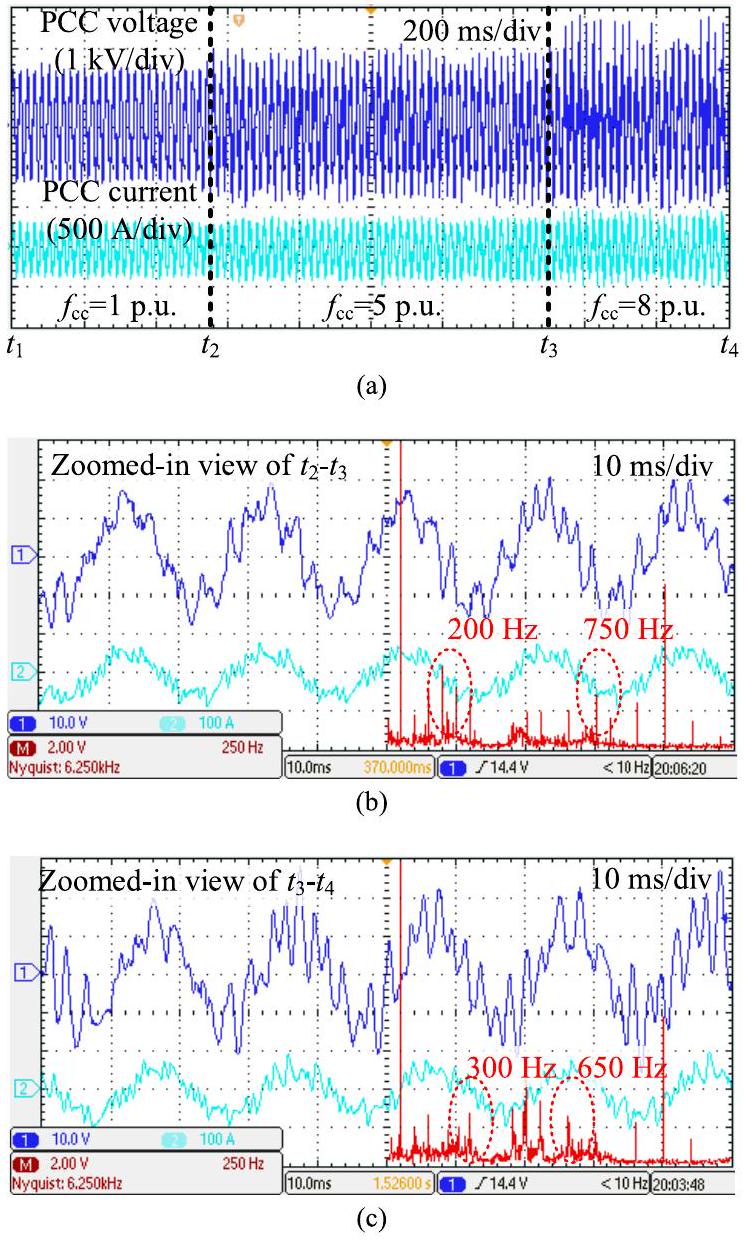}
\caption{Voltage and current waveforms under different control bandwidths. (a) $f_{\mathrm{cc}}$ increases from 1 to 5 p.u. at $t_{2}$, and then increases from 5 to 8 p.u. at $t_{3}$. (b) Zoomed-in view of period $t_{2}$ and $t_{3}$. (c) Zoomed-in view of period $t_{3}$ and $t_{4}$.}
\end{center}
\end{figure}

Fig. 15(a) compares the frequency responses under different control bandwidths. When the bandwidth is set to 1 p.u., the traction network admittance $Y_{\text{net }}(s)$ intersects with the locomotive input-admittance $Y_{\text{siso }}(s)$ at two points. The frequencies of these points are 281 and 449 Hz , respectively, with PDs of $68.2^{\circ}$ and $-118^{\circ}$, showing that the system is stable. When the bandwidth is increased to 5 p.u., the interaction points are locate at 278 and 585 Hz , with PDs of $116.3^{\circ}$ and $-116.7^{\circ}$. It is shown that as the bandwidth increases, the PM of the first intersection point decreases from $111.8^{\circ}$ to $63.7^{\circ}$, which means that the harmonic is more prone to be amplified around

\begin{figure}[H]
\begin{center}
  \includegraphics[alt={},max width=\columnwidth]{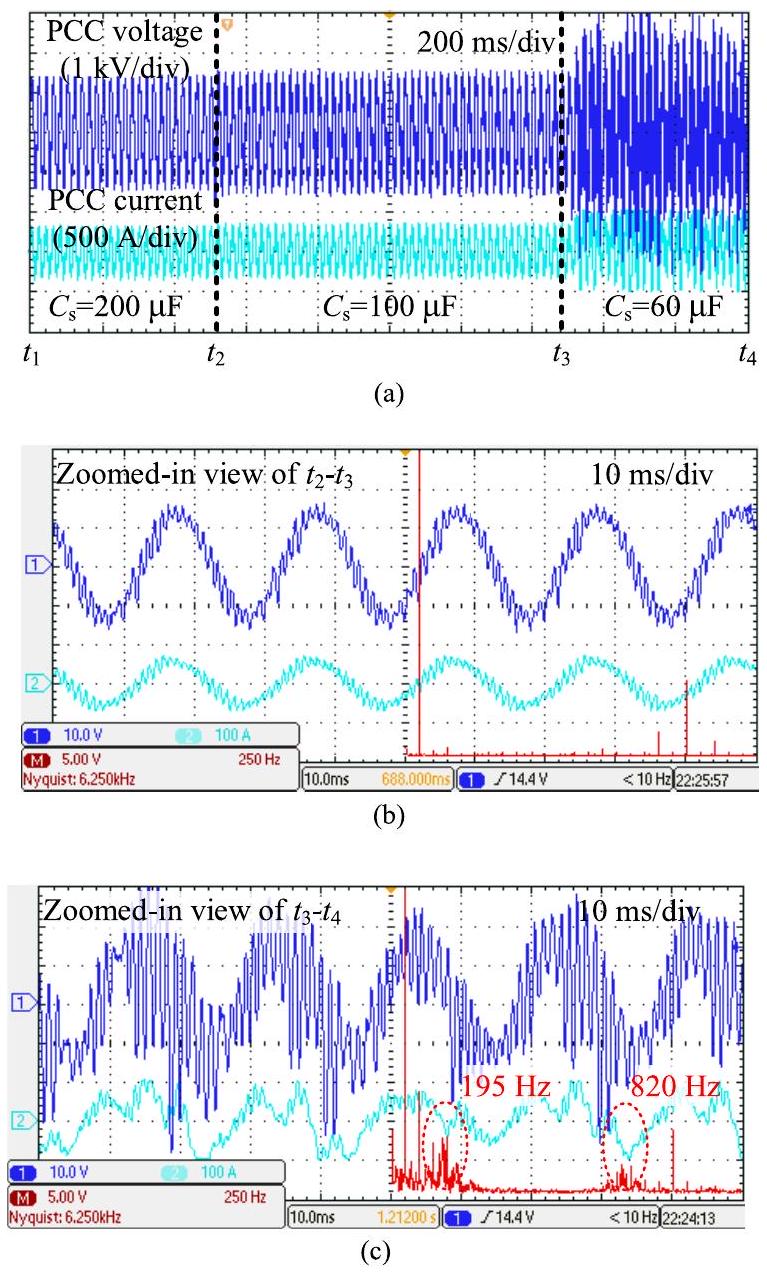}
\caption{Voltage and current waveforms under different traction network impedance. (a) $C_{s}$ decreases from 200 to $100 \mu \mathrm{~F}$ at $t_{2}$, and then increases from 100 to $60 \mu \mathrm{~F}$ at $t_{3}$. (b) Zoomed-in view of period $t_{2}$ and $t_{3}$. (c) Zoomedin view of period $t_{3}$ and $t_{4}$.}
\end{center}
\end{figure}

this frequency. Further, as the bandwidth increases to 8 p.u., $Y_{\text{siso }}(s)$ is less than $-90^{\circ}$ above 630 Hz , and $\mathrm{PM}<0$ at this time will directly cause high-frequency HIS in this frequency range.

Fig. 16 shows the voltage and current waveforms at PCC under different control bandwidths. In Fig. 16(a), it can be seen that the process includes a total of four time points from $t_{1}-t_{4}$, where the bandwidth is changed at $t_{2}$ and $t_{3}$. In $t_{1}$ and $t_{2}$, when the control bandwidth is set to 1 p.u., it is shown that the voltage and current are running steadily. In $t_{2}$ and $t_{3}$, when the bandwidth is set to 5 p.u. According to the\\
fast Fourier transform (FFT) analysis in Fig. 16(b), the harmonic will arise around 200 Hz , and the harmonic components at 750 Hz are the PWM sideband harmonic. Furthermore, the bandwidth is set to 5 p.u. in $t_{3}$ and $t_{4}$. According to Fig. 16(c), the generated harmonic components are mainly around 650 Hz , where the harmonic components at 300 Hz are the PWM sideband harmonic. Therefore, the HIL results confirm the effectiveness of the developed impedance model and the stability analysis results above.

Fig. 17 shows the voltage and current waveforms at PCC under different traction network impedance. In Fig. 17(a), the process similarly includes a total of four time points from $t_{1}-t_{4}$, where the equivalent capacitance is changed at $t_{2}$ and $t_{3}$. In $t_{1}$ and $t_{2}$, it is shown that the voltage and current are stable when the capacitance is set to $200 \mu \mathrm{~F}$. When the capacitance is set to $100 \mu \mathrm{~F}$ in $t_{2}$ and $t_{3}$, the voltage and current are stable without newly generated harmonics, according to the FFT analysis in Fig. 17(b). Then, in $t_{3}$ and $t_{4}$, the capacitance is decreased to $60 \mu \mathrm{~F}$. According to the FFT analysis in Fig. 17(c), the generated harmonic components are mainly around 820 Hz . In contrast, the harmonic components around 195 Hz are the PWM sideband harmonic. These generated harmonic frequencies correlate closely with the frequency of intersection points, indicating that the harmonics are caused by a marginally stable resonance between the locomotive and the traction network.

According to the theoretical analysis and HIL verification results, the established locomotive input-admittance model can accurately analyze the harmonic instability issues in the system. It is shown that the parameters variation will lead to different stability margins. In order to quantify and analyze the impact of different parameters on system stability, the sensitivity function analysis can be found in [22].

\section*{V. Conclusion}
This article reveals the dynamic propagations of input perturbation frequency and the generated sideband harmonic components in the PWM comparator, resulting in MIMO characteristics for locomotive input admittance. Then, considering the PWM sideband harmonic components, a multifrequency model is proposed to analyze the high-frequency performance of locomotive input admittance. It is found that the generated PWM sideband harmonic components dominate the locomotive input-admittance characteristics when the perturbation frequency is higher than the $1 / 2$ switching frequency. Afterward, a precise SISO model in high-frequency range is derived based on the conversion technique to simplify the stability analysis process. According to the SISO model, the influence of switching frequency, control bandwidth, and traction network impedance on system stability are revealed. The unstable harmonic components can be accurately identified by the stability analysis results.

\section*{References}
[1] K. Song, W. Mingli, S. Yang, Q. Liu, V. G. Agelidis, and G. Konstantinou, "High-order harmonic resonances in traction power supplies: A review based on railway operational data, measurements, and experience," IEEE Trans. Power Electron., vol. 35, no. 3, pp. 2501-2518, Mar. 2020.\\[0pt]
[2] H. Hu, Y. Shao, L. Tang, J. Ma, Z. He, and S. Gao, "Overview of harmonic and resonance in railway electrification systems," IEEE Trans. Ind. Appl., vol. 54, no. 5, pp. 5227-5245, Sep. 2018.\\[0pt]
[3] Railway Applications-Power Supply and Rolling Stock Technical Criteria for the Coordination Between Power Supply (Substation) and Rolling Stock to Achieve Interoperability," Cenelec Standard EN 50388 Ed.2, 2012.\\[0pt]
[4] J. Sun, "Impedance-based stability criterion for grid-connected inverters," IEEE Trans. Power Electron., vol. 26, no. 11, pp. 3075-3078, Nov. 2011.\\[0pt]
[5] S. Wu and Z. Liu, "Low-frequency stability analysis of vehiclegrid system with active power filter based on $d q$-frame impedance," IEEE Trans. Power Electron., vol. 36, no. 8, pp. 9027-9040, Aug. 2021.\\[0pt]
[6] Y. Qiu, M. Xu, K. Yao, J. Sun, and F. C. Lee, "Multifrequency small-signal model for buck and multiphase buck converters," IEEE Trans. Power Electron., vol. 21, no. 5, pp. 1185-1192, Sep. 2006.\\[0pt]
[7] Y. Qiu, M. Xu, J. Sun, and F. C. Lee, "A generic high-frequency model for the nonlinearities in buck converters," IEEE Trans. Power Electron., vol. 22, no. 5, pp. 1970-1977, Sep. 2007.\\[0pt]
[8] L. Harnefors, R. Finger, X. Wang, H. Bai, and F. Blaabjerg, "VSC input-admittance modeling and analysis above the Nyquist frequency for passivity-based stability assessment," IEEE Trans. Ind. Electron., vol. 64, no. 8, pp. 6362-6370, Aug. 2017.\\[0pt]
[9] F. D. Freijedo, M. Ferrer, and D. Dujic, "Multivariable high-frequency input-admittance of grid-connected converters: Modeling, validation, and implications on stability," IEEE Trans. Ind. Electron., vol. 66, no. 8, pp. 6505-6515, Aug. 2019.\\[0pt]
[10] D. Yang, X. Wang, and F. Blaabjerg, "Sideband harmonic instability of paralleled inverters with asynchronous carriers," IEEE Trans. Power Electron., vol. 33, no. 6, pp. 4571-4577, Jun. 2018.\\[0pt]
[11] H. Tao, H. Hu, X. Zhu, K. Lei, and Z. He, "A multifrequency model of electric locomotive for high-frequency instability assessment," IEEE Trans. Transport. Electrific., vol. 6, no. 1, pp. 241-256, Mar. 2020.\\[0pt]
[12] H. Y. Assefa, S. Danielsen, and M. Molinas, "Impact of PWM switching on modeling of low frequency power oscillation in electrical rail vehicle," in Proc. 13th Eur. Conf. Power Electron. Appl., Barcelona, Spain, Sep. 2009, pp. 1-9.\\[0pt]
[13] X. Yue, F. Zhuo, S. Yang, Y. Pei, and H. Yi, "A matrix-based multifrequency output impedance model for beat frequency oscillation analysis in distributed power systems," IEEE J. Emerg. Sel. Topics Power Electron., vol. 4, no. 1, pp. 80-92, Mar. 2016.\\[0pt]
[14] Z. Liu, G. Zhang, and Y. Liao, "Stability research of high-speed railway EMUs and traction network cascade system considering impedance matching," IEEE Trans. Ind. Appl., vol. 52, no. 5, pp. 4315-4326, Sep. 2016.\\[0pt]
[15] Y. Song, Z. Wang, Z. Liu, and R. Wang, "A spatial coupling model to study dynamic performance of pantograph-catenary with vehicletrack excitation," Mech. Syst. Signal Process., vol. 151, Apr. 2021, Art. no. 107336.\\[0pt]
[16] Y. Song, Z. Liu, A. Rønnquist, P. Navik, and Z. Liu, "Contact wire irregularity stochastics and effect on high-speed railway pantograph-catenary interactions," IEEE Trans. Instrum. Meas., vol. 69, no. 10, pp. 8196-8206, Oct. 2020.\\[0pt]
[17] Q. Ma, "Study on some problems of traction power supply systems in high-speed railway," Ph.D. dissertation, School Elect. Eng., Southwest Jiaotong Univ., Chengdu, China, 2013.\\[0pt]
[18] D. G. Holmes and T. A. Lipo, Pulse Width Modulation for Power Converters: Principles and Practice. Hoboken, NJ, USA: Wiley, 2003.\\[0pt]
[19] H. D. T. Mouton, B. McGrath, D. G. Holmes, and R. H. Wilkinson, "One-dimensional spectral analysis of complex PWM waveforms using superposition," IEEE Trans. Power Electron., vol. 29, no. 12, pp. 6762-6778, Dec. 2014.\\[0pt]
[20] M. Sowa, "A harmonic balance methodology for circuits with fractional and nonlinear elements," Circuits, Syst., Signal Process., vol. 37, no. 11, pp. 4695-4727, Nov. 2018.\\[0pt]
[21] C. Zhang, M. Molinas, S. Foyen, J. A. Suul, and T. Isobe, "Harmonicdomain SISO equivalent impedance modeling and stability analysis of a single-phase grid-connected VSC," IEEE Trans. Power Electron., vol. 35, no. 9, pp. 9770-9783, Sep. 2020.\\[0pt]
[22] H. Hu, H. Tao, X. Wang, F. Blaabjerg, Z. He, and S. Gao, "Train-network interactions and stability evaluation in high-speed railways-Part II: Influential factors and verifications," IEEE Trans. Power Electron., vol. 33, no. 6, pp. 4643-4659, Jun. 2018.\\

\end{document}